# Photo Screen: Shaping Perceptions of Residential Communities

Holger Schnädelbach, Tom Lodge
University of Nottingham
Nottingham, UK
pszhms, psztl @nottingham.ac.uk

Tim Coughlan
The Open University
Milton Keynes, UK
Tim.Coughlan@open.ac.uk

Alex Taylor
City University of London
London, UK
Alex.Taylor@city.ac.uk

**ABSTRACT**
Engaging residential communities with each other and with management remains a challenge. Housing providers deploy a variety of engagement strategies, some of which are supported by digital technologies. Their individual success is varied and integrated, multi-pronged approaches are seen to be more successful. As part of those, it is important to address people's perceptions of community and places, as well as any practical issues that they face. We present the design and evaluation of Photo Screen, a situated, public photo taking and viewing screen which was deployed in the context of a new flagship housing estate as part of a range of community engagement measures. In a new context, we confirm the high levels of engagement that can be achieved with this simple mechanism. We propose that photo 'tagging' might offer a second-stage engagement mechanism and enable meaningful dialogue between residents and management. Finally, we discuss how this playful activity allowed residents to positively shape the perception of their community.

**ACM Classification Keywords**
H.5.1 [Multimedia Information Systems]; H.5.3 [Group and Organization Interfaces]

**INTRODUCTION**
Residential, place-based communities typically develop over extended periods of time and their 'success' depends on a whole host of factors related to the physical and social characteristics of a particular neighbourhood. Longstanding work has established the importance of satisfaction measures to determine whether people are committed to a particular community, and contribute to its long-term development [1]. Adriaanse has argued that satisfaction with the 'social climate', i.e. the community aspects of a place beyond the physical properties of a location, is a key factor in whether someone is satisfied with their surroundings [2].

Research then points to how people gain satisfaction and commitment to their residential estates and neighbourhoods when they feel bound to the place, to not just its physical but also social geography. However, this line of research becomes harder to reconcile as distinctions are made between objective and subjective measures of satisfaction [3], and perceptions are understood to be somehow separate from normative measures (e.g. [4]). Absent in these accounts seem to be any recognition of the active and unfolding practices people engage in to enact or do community [5]. More specifically, it's hard to see how the kinds of technological interventions we are familiar with in HCI make an actual difference to civic life and not just people's subjective or perceived experiences (see [6]).

The use of online forums or social media groups for residential communities presents an example. Both—regularly put in place by management to facilitate the discussion of issues—are often vulnerable to uninhibited behaviour and negative sentiment. These issues can be tied to the particular features of the forums and social media: a lack of easy expression of social cues and non-verbal feedback [7, 8], the affordances of certain structures (e.g. anonymity, moderation) and the quality of content (e.g. political/personal views) [9, 10]. Technologies like forums are, in short, bound up with not just how people subjectively perceive their environment, but how they actually live together.

Here we report on a study of a system that has been designed to aid this active doing of community: *Photo Screen*. The system was developed, amongst a range of other measures, to allow residents to create and share photos with others they live with. We review the development context, the system itself, and related work in HCI. This is followed by the analysis of photos taken with Photo Screen and its related data. We conclude by reflecting on the value of a tool that, as we'll show, gives people opportunities to visibly perform residential life in ways that are highly positive and demonstrably communal.

**Housing association collaboration and background**
The work we present stems from an extended collaboration with a large UK housing association (HA) and its investment in a multimillion-pound regeneration programme targeting one of the most stigmatized and unpopular, ex-council estates in Greater London. Around 600 homes in five fourteen story blocks were to be replaced with the same amount of low-rise homes, moving away from mostly 1-bed flats to a mix of housing units and ownership models. The ambition of the HA's project manager was to transform the neighbourhood into the most desirable in the area, without creating a '*yuppie enclave*' and she summarised her aim as *'...building communities and providing great homes for people who need them.'*. This was in the context of well-documented difficulties that communities face when such radical redevelopment is undertaken (e.g. [11] [12]).

Recognising the inherent troubles of top-down 'community engineering', the HA focussed on practical, on-the-ground measures to help *'... kick-start them a bit'.'* This involved inviting people from the old estate to document history, co-develop public art and welcome newcomers. The HA employed a community support officer to help with the transition, encouraged blogging and a Facebook group and it continues to involve residents in the new development. Beyond these existing measures, the HA was keen to *'...widen our repertoire and find new ways of building sustainable community networks using technology and the arts.'*

**Photo Screen Deployment**
We discussed a host of applications (e.g. games, message boards, questionnaires) for a screen deployment. Our previous experience with a public photo-taking application [21] had resulted in very high engagement levels. Focusing on its key design principle we iterated development in collaboration with the HA (e.g. layout, tags and management messages added). Photo Screen has the following four functions. It allows people to: 1) Take photos, 2) Tag photos from a set of eight pre-defined tags, 3) View recent photos and tags and 4) management can post messages on screen.

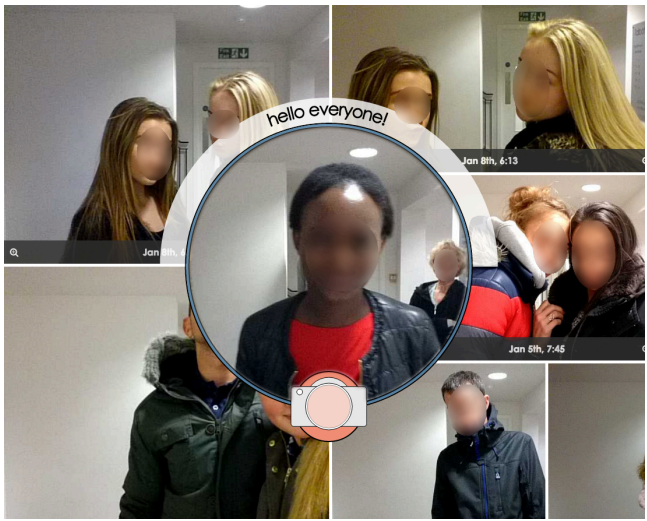

**Figure 1: Main screen with live viewfinder in centre. Tapping screen starts countdown to photo being taken (see Fig. 2). Viewfinder is framed by messages selected by management and the latest photos. Tapping any photo displays it with tags.**

Photo Screen was deployed on a standard 10 inch Android tablet. The application was developed as a webapp running in a full-screen Chrome browser. We used a kiosking app to lock down interaction to the application only. The tablet was securely mounted in the foyer of one of the three new residential apartment blocks.

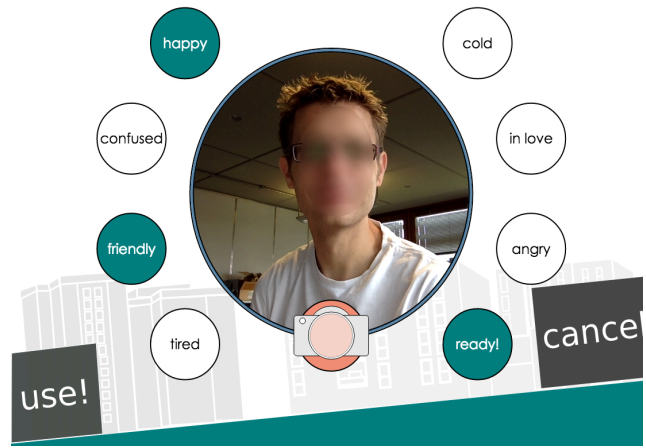

**Figure 2 Following countdown, the captured photo is shown. Photos can be used or cancelled. The figure shows three of eight possible tags being selected.**

The block has a total of 25 apartments (one and two bedrooms). At the time of the deployment, there were 61 occupants (19 males, 23 females and 19 children). The tablet's front camera overlooked the corridor that everyone entering the building passed, and it was facing the lift entrance. This location was selected to maximise the potential for engagement. Access to the foyer was controlled by a keypad for the block entrance. Access to the photo screen was open to anyone present in the foyer. The screen was labelled with the following instructions: *'This is for you to use and try out, all you have to do is press the camera button in the middle of the screen. It takes a selfie within 5 seconds and you can add a description of how you are feeling. If you don't like the photo, press cancel and try again.'* The eight optional tags were also printed on signs either side of the screen. In addition, we displayed information about data capture and retention and contact details for reporting concerns. As this was a semi-public space, we expected no or fewer problematic images compared to more public settings [13]. No reports of problematic images were made and we did not have to use the post-hoc moderation feature.

In summary, the photo screen was very much public within the closed community of the block and highly visible. It was designed to be simple to use and to encourage play.

**Related work**
A variety of interactive prototypes for community engagement have been developed and evaluated by the HCI community. The Tenison Road project surfaced community generated data on the street [14], making community activities more visible. Neighborland provided a platform for people to discuss improvements to their surroundings [15]. In the Livehoods project, social media generated by residents was used to represent areas of the city [16], allowing people to implicitly generate a representation of their surroundings. There are also a host of projects that involve the installation of public screens in various settings. The Dynamo communal display allowed people to share media, mirroring and extending existing social practices [17, 18]. In urban space, the most extensive screen network has been established in Oulu, emphasising the potential for two-way communication [20]. This network included public photo taking to engage youngsters in expressing their opinions [21]. Closest to the photo screen reported here is the open-ended system developed for the SITW network [13] (from which we draw our evaluation methodology) [19]. We are not aware of any similar photo taking application having been deployed or evaluated in support of the development of closed residential communities.

| Group Portrait | Single Portrait | Facial Expressions | Affection | Action |
|---|---|---|---|---|
| (81) **42%** | (69) **36%** | (32) **17%** | (28) **15%** | (25) **13%** |
| 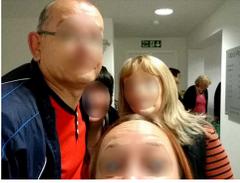 | 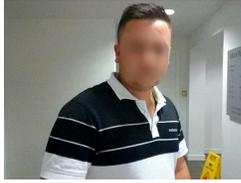 | 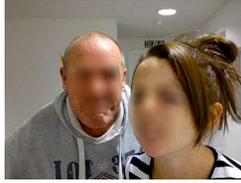 | 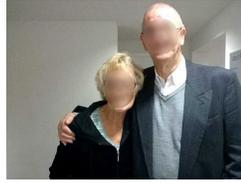 | 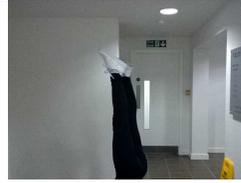 |

**Figure 3** The five photo categorisations (see [19]) represented most frequently in the Photo Screen data

**Photo screen in use**

We handed the prototype to the HA, who managed its availability, and we observed the deployment in late 2015. The networking on location was unreliable and every runtime was enabled by the community support officer. Despite this, photo screen was up for 18 days of the 49-day deployment period, mostly during daytime hours and this time is sufficient to draw the presented conclusions. Insights into the photo screen system were assembled from: analysis of the photos taken (e.g. frequencies and content); a brief feedback session with community members; a survey with nine residents (all who made themselves available; 15% of block residents) asking about the functionality of the system and motivations for using it (in multiple-choice and open-ended questions); and on-going discussions with the estate's community team.

*Photos taken*

192 photos were taken. Using a standard operating system face recognition feature and input from the HA, 54 individuals were identified in the photos. Discounting trials and demonstrations of the system, the photos showed 24

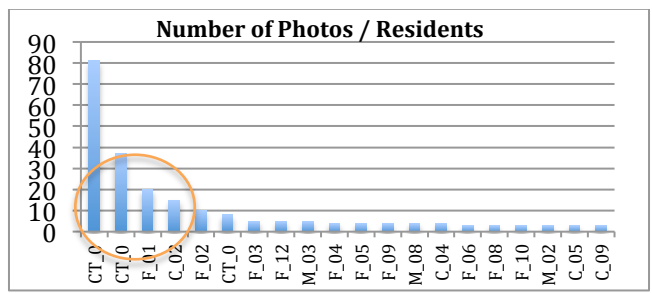

**Figure 4** Number of appearances against identified individuals, plotting all individuals who appear three times or more.

**F=Female adult, M=Male adult, CT=Child Teenager (judged to be taking photos by themselves), C=Child (often held into camera and judged not to take photos themselves).**

females, 17 males and 13 children, roughly mirroring the demographics of the block. Thirty-six of the 50 individuals captured lived in the apartment block where the screen was installed, still amounting to ~60% of the block's residents. It is common for individuals and groups to dominate public screens like this. One teenager appeared in ~42% (81) of all photos and another in ~19% (37), while 31 people appeared only in one or two photos. Fig. 3 illustrates how photo taking was further influenced by people knowing each other, with the four most active users frequently appearing in photos together (CT_02, CT_01, F_01 and C_02).

*Photo categories*

We applied the method presented in [19] to analyse the photos further. This was done by a single researcher in a first round and then spot-verified by a second researcher. The five most popular categories were Group Portrait ([19] label: *A group of people having a portrait photo*), assigned to 42% of photos and Single Portrait (*A person having a portrait photo facing the camera frontally*) assigned to 36%. Facial Expressions (*Having a facial expression in the photo, e.g., lolling/showing tongue out, making a "silly" face, or duck mouth*), Affection (*People showing affection to each other, for example kissing or hugging each other*) and Action (*A sequence of photos catching people in "action" (where we also included single photos classified showing people in 'action', for example making a handstand*), were found in 13% to 17% of photos (compare Fig. 2). The other five categories were represented much less.

Compared to [19], from which we borrow the categorisation scheme, it is striking that, in this context, the Photo Screen produced twice the proportion of photos categorised as 'Group Portrait and ten times the proportion of photos categorised as 'Affection'. It is also worth noting that no photo was categorised as 'Inappropriate'.

In the survey interview, four residents listed 'Fun' as the reason for why they left photos on the screen, and one child stating: '*Great fun. Loved looking at everyone who took pictures.*' Others had more specific aims, with one resident wanting '*...to get to know people*' and someone else saying they '*...wanted it to appear friendly to others.*'.

*Photos viewed*

The system was as much a community display as it was a camera, allowing residents to view the most recent photos taken, tags and management messages. All survey respondents stated that they looked at others' photos as well as their own, suggesting the communal quality of the display was appealing.

*Photo Tagging*

The interface allowed photos to be tagged with a selection from 8 words and up to 8 times. Tags represented a broad range of feelings relevant for the situation of taking of photo of one-self. Free tagging would have a required an input method, which would have broken the immediacy of the interaction. Of the 192 photos taken, 152 were tagged at least once, i.e., nearly 80% of all photos were tagged. Table 1 shows all tags and their frequency.

**Table 1 Tags and tag frequency**

| **Happy** | **Friendly** | **Confused** | **Ready!** |
|---|---|---|---|
| (48) **25%** | (39) **20%** | (29) **15%** | (27) **14%** |
| **Tired** | **In Love** | **Cold** | **Angry** |
| (27) **14%** | (24) **12%** | (13) **7%** | (9) **5%** |

The two most frequent tags used were happy and friendly with the least frequent tags being cold and angry, demonstrating a preference for tags portraying a more positive outlook of the photos shown. The high proportion of photos tagged points to the fact that this feature was easy to use alongside taking photos. However, only three of the surveyed residents stated that they looked at any of the tags.

*Community messages*

Management used the message feature, posting 8 different messages in the study period. The first message was '*Welcome your neighbours*', followed by a series of event announcements, e.g. '*Quiz night this Thursday at 7pm*' and simple statements such as '*Have a great weekend*'. The photos captured by management included two leaflets held to camera, for additional detail. Despite value being put on messaging by management, only two survey respondents confirmed that they looked at the messages.

*Getting to know people*

Survey respondents were divided about whether the photo screen helped them get to know others living in the block. Two people reported that they got to know people because of the system, one of those expanding that it helped *'recognising who lived and visited the block'*. Less favourably, one resident stated *'[I] didn't find it helped me in anyway to recognise my neighbours. So it didn't work for me.'* Two people put a priority on physically meeting others, for example via activities put on by the housing association or meeting fellow residents at the lift entrance.

*Management feedback*

Management reported that the Photo Screen proved highly popular and, unexpectedly, interest was sustained over a long period of time. Management valued the impact that a small degree of interactivity had on resident engagement. For practical reasons (installation cost and monitoring), management is aiming to build on the lessons learnt in the deployment of a single, larger screen centrally on site.

**DISCUSSION**

There was sustained and repeated interaction with the system: residents took and tagged photos, and regularly engaged with others' photos. The work presented here confirms how this simple mechanism of communal photo taking results in very high levels of engagement also in this new context of residential communities (previously reported in an urban context (19)). <u>Lesson 1</u>: *Designers should consider this mechanism in other community engagement contexts to get people over the initial engagement thresholds.* Related to this, we found that residents did not value direct messages from management displayed on the same screen. However, tagging photos was surprisingly popular, despite it generating extra 'work' beyond the more playful engagement. <u>Lesson 2:</u> *We propose that an adapted 'photo tagging' activity that presents a space to comment, encourage, complain and discuss alongside community photos will be a valuable second-stage mechanism to engage residents.* Finally, reflecting on the wider context on of this work, we want to emphasise the positive role of systems like Photo Screen in residential estates and specifically ones with a troubled history. There was a quality to Photo Screen's content that contested the statistics of crime and depravation that had marked the estate. Because of its simple, walk-up interaction, photo screen gave residents (individuals and groups) the opportunity to present themselves to others. These individual acts in the moment, then amassed to reflexively create an image of this particular community and its presentation back to residents. The happy group portraits and affection shown in them, the positive tags, and the lack of inappropriate photos, all revealed a kind of work being put into performing collective life that contrasted with the estate's history, and also with the negative content found in many community forums. <u>Lesson 3</u>: *We need to consider how to draw focus away from building systems that project community onto a place or assert a model of satisfaction from outside. Rather, it appears more fruitful to think about how systems enable people to enact or perform their communities, and how such systems are part of a larger ecology of approaches with the same aim.*

**Acknowledgements**

We gratefully acknowledge support from Orbit Housing association and especially Caroline Field and Francois Jensen. This research was supported by EPSRC grant EP/G065802/1 and by the University of Nottingham through the Nottingham Research Fellowship 'The Built Environment as Interface to Personal Data'. We also thank all Photo Screen participants.